\documentclass[aps, pra, preprint, floatfix, amsmath, superscriptaddress]{revtex4-1}

\usepackage[separate-uncertainty = true,multi-part-units=single]{siunitx}	% SI unit symbols
\usepackage{bm}
\usepackage{graphicx}
\usepackage{amssymb}
\usepackage{amsfonts}
\usepackage{xcolor}
\usepackage[labelfont=bf]{caption}
\usepackage{hyperref}

\begin{document}

\title{The visual appearances of disordered optical metasurfaces}

\author{Kevin Vynck}
\email{kevin.vynck@univ-lyon1.fr}
\altaffiliation{Current address: Univ. Claude Bernard Lyon 1, CNRS, iLM, F-69622 Villeurbanne, France}
\affiliation{Univ. Bordeaux, IOGS, CNRS, LP2N, F-33400 Talence, France}

\author{Romain Pacanowski}
\affiliation{Univ. Bordeaux, IOGS, CNRS, LP2N, F-33400 Talence, France}
\affiliation{INRIA Bordeaux Sud-Ouest, F-33400 Talence, France}

\author{Adrian Agreda}
\affiliation{Univ. Bordeaux, IOGS, CNRS, LP2N, F-33400 Talence, France}

\author{Arthur Dufay}
\affiliation{INRIA Bordeaux Sud-Ouest, F-33400 Talence, France}

\author{Xavier Granier}
\affiliation{Univ. Bordeaux, IOGS, CNRS, LP2N, F-33400 Talence, France}

\author{Philippe Lalanne}
\email{philippe.lalanne@institutoptique.fr}
\affiliation{Univ. Bordeaux, IOGS, CNRS, LP2N, F-33400 Talence, France}

\begin{abstract}

Nanostructured materials have recently emerged as a promising approach for material appearance design. Research has mainly focused on creating structural colours by wave interference, leaving aside other important aspects that constitute the visual appearance of an object, such as the respective weight of specular and diffuse reflectances, object macroscopic shape, illumination and viewing conditions. Here, we report the potential of disordered optical metasurfaces for harnessing visual appearance. We develop a multiscale modelling platform for the predictive rendering of macroscopic objects covered by metasurfaces in realistic settings, showing how nanoscale resonances and mesoscale interferences can be used to spectrally and angularly shape reflected light and thus create unusual visual effects at the macroscale. We validate this property with realistic synthetic images of macroscopic objects and centimetre-scale samples observable with the naked eye. This framework opens new perspectives in many branches of fine and applied visual arts.

\end{abstract}

\maketitle

%%%%%%%%%%%%%%%%%%% INTRODUCTION %%%%%%%%%%%%%%%%%%%

Many beautiful appearances in the living world are created by wave interference in micro or nanostructured materials~\cite{vukusic2003photonic, kinoshita2008structural}. Drawing inspiration from nature, a broad variety of iridescent and diffuse structural colours have now been obtained with artificial materials, including low-index chiral nematic architectures~\cite{shopsowitz2010free, parker2018self, chan2019visual}, liquids of dielectric particles~\cite{takeoka2009structural, park2014full, goerlitzer2018bioinspired} and more recently, microscale droplets~\cite{goodling2019colouration}.

Finely-engineered high-index resonant scatterers methodologically arrayed on a substrate offer an alternative approach. Such ``optical metasurfaces'' have attracted considerable attention recently for various applications~\cite{yu2014flat, arbabi2015dielectric, kuznetsov2016optically, Chavel2017}, including the realization of ultra-thin multicoloured surfaces for high-resolution displays, wherein the resonant scatterers act as nanopixels with versatile chromatic properties~\cite{zhu2017resonant, kristensen2017plasmonic, stewart2017toward, peng2019scalable, daqiqeh2020nanophotonic}. The potential of optical metasurfaces for the design of appearance has however remained quite unexplored due to the difficulty to relate optical properties at the nanoscale to the manifold attributes of appearance at the macroscale. Visual appearance is indeed much more than just colour. For instance, gloss and haze, which rely on the interplay between the diffuse and specular components of the scattered light, are as essential as colour for our perception of objects~\cite{hunter1987measurement, fleming2003real}. The appearance of an object also tightly depends on its macroscopic shape, as well as on the lighting and viewing conditions -- iridescent effects are for instance better appreciated when the object or the observer move. All these aspects have not been considered in previous studies on optical metasurfaces, and incompletely investigated in most works on natural and artificial structural colours.

Appearance attributes and their role for perception have been thoroughly studied by the computer graphics community, which has developed advanced tools to generate synthetic images with an exceptionally high realism~\cite{pharr2016physically}. Rendering relies on ray tracing algorithms to propagate light in virtual environments and accurately models how light is reflected from material surfaces thanks to a high-dimensional function known as the bidirectional reflectance distribution function (BRDF)~\cite{nicodemus1977geometrical}. Wave-optics BRDF models have been developed to render a large set of surface finishes featuring iridescent and diffraction effects~\cite{stam1999diffraction, cuypers2012reflectance, musbach2013full, holzschuch2017two, belcour2017practical, werner2017scratch} or to synthesize custom reflectances~\cite{weyrich2009fabricating, levin2013fabricating}. However, due to the unavailability of dedicated BRDF models~\cite{tsang2004scattering}, the visual appearance of optical metasurfaces featuring both specular and diffuse components has not been investigated yet. 

Here, we show how subwavelength-scale interferences in disordered optical metasurfaces can be engineered to create impressive visual appearances. Resonant scattering at the nanoscale~\cite{kuznetsov2016optically}, multiple scattering at the mesoscale~\cite{tsang2004scattering} and light transport at the macroscale~\cite{pharr2016physically} are combined into a multiscale modelling platform [Fig.~\ref{fig1}a-c] to unravel how nanostructure features translate into distinct visual effects. We demonstrate the possibility to control quasi-independently the colours of the specular and diffuse components and report two visual effects that are particularly impressive when changing viewing conditions and that cannot be obtained with alternative strategies like optical multilayers~\cite{maile2005effect}, low-index photonic materials~\cite{goerlitzer2018bioinspired} or microscale textures~\cite{weyrich2009fabricating, levin2013fabricating}. One of these effects is demonstrated with centimetre-scale metasurfaces observable by the naked eye under various illumination and viewing conditions.

\section*{Predictive rendering of disordered metasurfaces}

\begin{figure*}[h!]
	\centering
	\includegraphics[width=\textwidth]{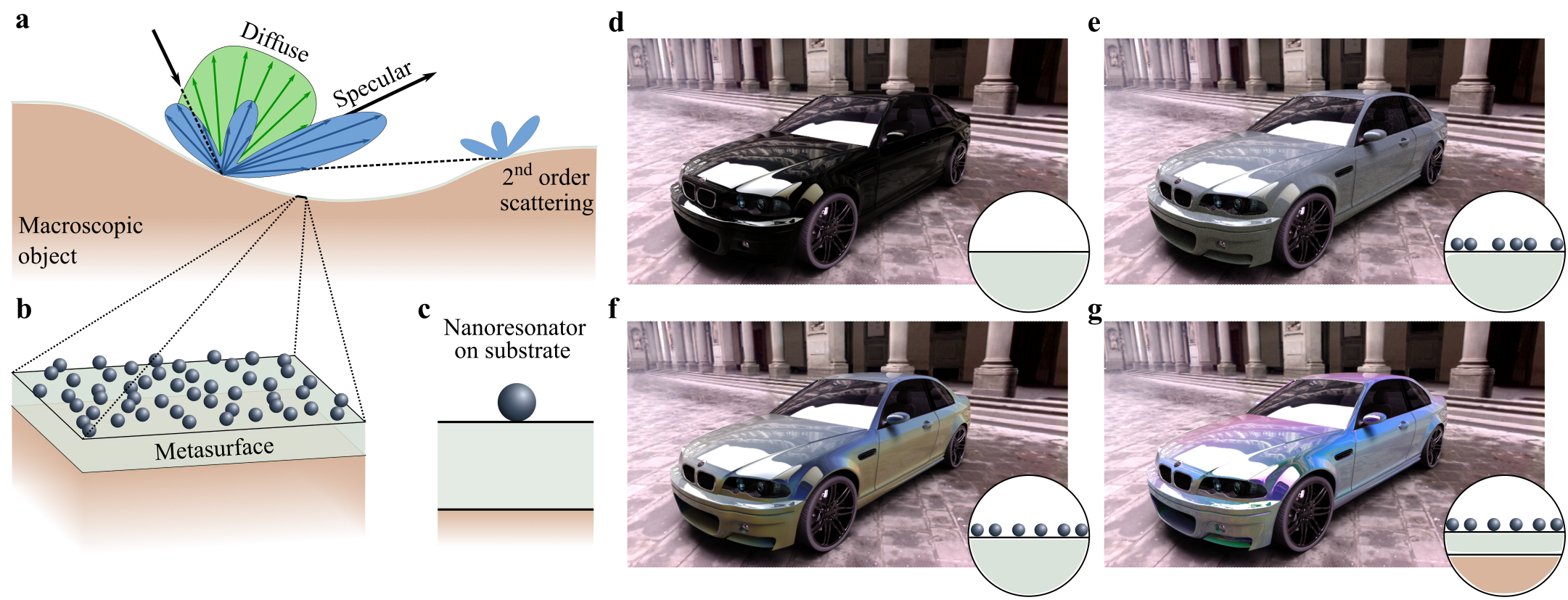}
	\caption{\textbf{Prediction of visual appearance of macroscopic objects covered by disordered metasurfaces.} \textbf{a}-\textbf{c}, Multiscale nature of the visual appearance of disordered metasurfaces, from the individual particle resonantly interacting with the substrate interfaces (\textbf{c}) to the ensemble of particles interacting together to form collective scattering effects (\textbf{b}), to the macroscopic surface that can experience multiple internal reflections (\textbf{a}). \textbf{d}-\textbf{g}, Synthetic images, obtained with our multiscale modelling platform, of a macroscopic object (here, a car) in a realistic lighting environment (here, in front of the Uffizi gallery in Florence, Italy). In \textbf{d}, the car body is made of an unstructured tinted glass, leading to a purely specular and dominantly black appearance. In \textbf{e}, the same surface is covered by Ag particles of radius $r=90$ nm distributed at random positions with a surface coverage $f=0.1$. Scattering by the particles reduces specular reflection and yields a diffuse, nearly isotropic, grey appearance. In \textbf{f}, a control of structural correlations (described by the parameter $p=0.5$, defined below) in the particle position leads to the emergence of a striking new appearance. In \textbf{g}, a structured substrate (SiO$_2$ layer of thickness $h=400$ nm on a Si substrate) is used, considerably enriching the observed diffuse colours.}
	\label{fig1}
\end{figure*}

The key role played by subwavelength-scale interferences on visual appearance is demonstrated in Fig.~\ref{fig1}, where we simulate the appearance of a familiar object (here, a car) covered by disordered metasurfaces in a realistic lighting environment (here, in front of the Uffizi gallery in Florence, Italy), see the Methods for details on modelling. With an unstructured absorbing substrate, the object has a common specular, dominantly black appearance. By adding scattering silver (Ag) particles at random positions on the surface, the specular reflection is reduced and the object takes a nearly uniform diffuse grey colour due to the broadband particle resonance. More interestingly, upon adding a short-range correlation in the particle positions, strictly the same object offers a very different visual appearance with vivid colours appearing out of the grey. The visual appearance is further enriched considerably when using a layered substrate, yielding an uncommon mix of blue, pink, green and violet. The visual appearances shown in the last two images strikingly differ from the uniform colours that would be expected from disordered photonic materials made of resonant particles~\cite{goerlitzer2018bioinspired, schertel2019structural} due to these additional interference phenomena. As we will show below, they also exhibit unique dynamic properties that appear when the lighting environment or the viewpoint changes.

%%%%%%%%%%%%%%%%%%% BRDF model %%%%%%%%%%%%%%%%%%%

The key element to simulate and understand the origin of these exotic visual appearances is our BRDF model, which has the great benefit of disentangling the respective roles of the individual particles and their spatial arrangement on reflectance. The BRDF of disordered materials is generally decomposed into a specular (coherent) component and a diffuse (incoherent) component, whose contributions drive the glossiness or mattness of the surface~\cite{hunter1937methods, elfouhaily2004critical}. Formally, the specular and diffuse components are respectively related to the average electromagnetic field produced by the surface and to the field fluctuating around the average~\cite{tsang2004scattering, elfouhaily2004critical}. Whereas advanced multiple-scattering models have been successfully developed for the specular reflectance and transmittance by monolayers of particles on substrates~\cite{hong1980multiple, garcia2012multiple}, state-of-the-art models for the diffuse component apply either to particle monolayers suspended in a uniform medium~\cite{loiko2018incoherent} or to particles much smaller than the wavelength~\cite{sasihithlu2016surface}, which is insufficient for our purpose. The major difficulty is to account for the effect of neighboring particles on the scattering properties of a resonant particle placed on a substrate. We overcome this issue here by assuming that the particles are excited by an average field (i.e., using a mean-field approximation~\cite{tsang2004scattering}) whose amplitude is determined from the specular reflectance and transmittance of the particle monolayer on the substrate, see the Supplementary Note 1 for technical details. The BRDF diffuse component [Eq.~\eqref{eq:BRDF-dif} in Methods] then takes the form of the expression obtained in the independent scattering approximation, yet including an important heuristic correction coefficient that ensures a more physically sound treatment, leading to quantitative predictions at high particle densities and grazing incident and scattered angles. Series of full-wave multiple-scattering computations on ensembles of particles reported in the Supplementary Note 2 fully validate the model, showing that it captures very well the main spectral and angular features of the reflected intensity for all situations considered in the present work.

%%%%%%%%%%%%%%%%%%% VISUAL APPEARANCE CONTROL %%%%%%%%%%%%%%%%%%%

\section*{Impact of metasurface parameters on visual appearance}

\begin{figure*}[h!]
  \centering
  \includegraphics[width=\textwidth]{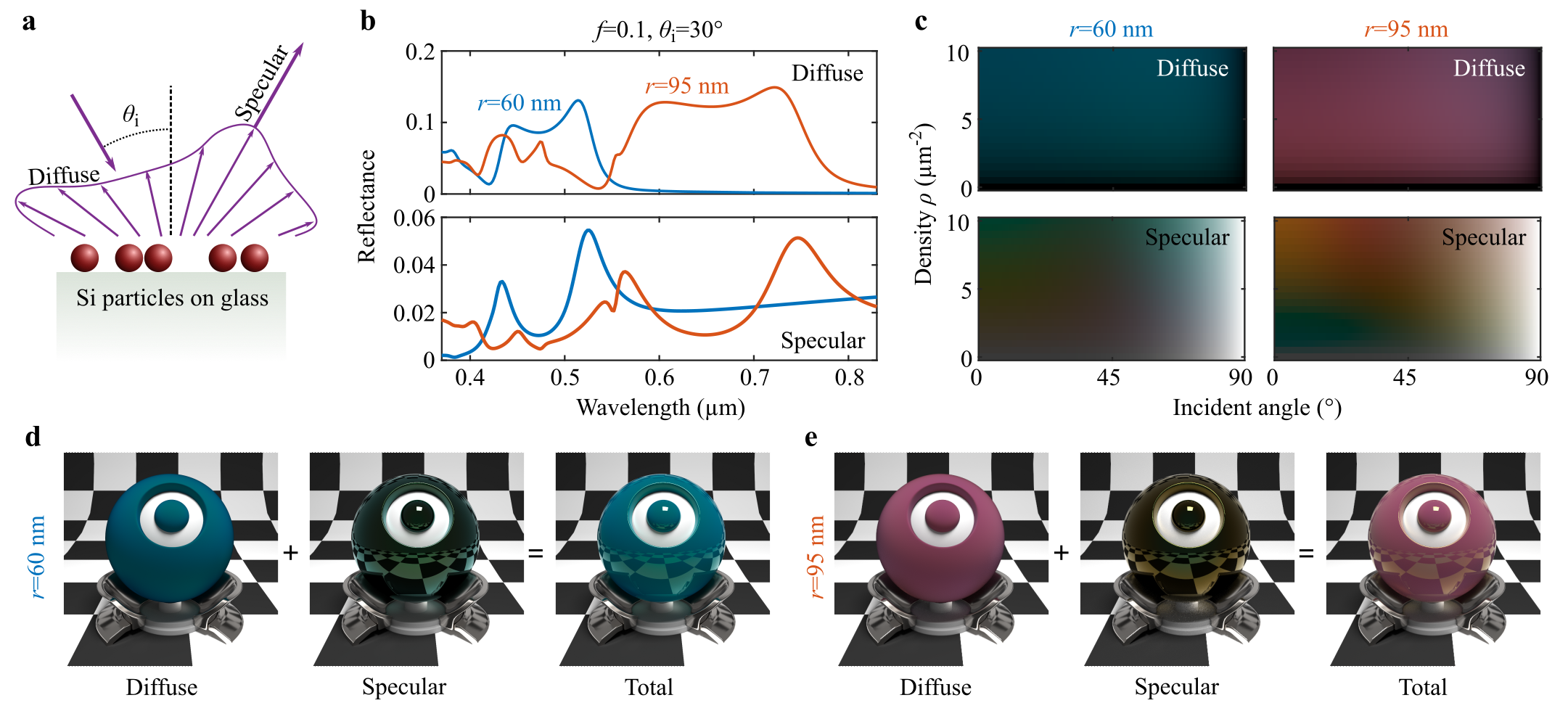}
  \caption{\textbf{Engineering of individual particles.} \textbf{a}, Spherical Si particles are placed randomly on a glass substrate. \textbf{b}, Spectra of the diffuse (top) and specular (bottom) reflectances of the metasurfaces predicted from the BRDF model for particle radii $r=60$ nm and $r=95$ nm at a surface coverage $f=0.1$, corresponding to particle densities $\rho=8.84$ $\mu$m$^{-2}$ and $\rho=3.53$ $\mu$m$^{-2}$, respectively. The incident light is unpolarised and makes an angle $\theta_\text{i} = 30^\circ$ with respect to the surface normal. The diffuse reflectance is integrated over the upper hemisphere. Changing the particle size shifts the particle resonances, as expected. The spectra of the diffuse and specular reflectances are markedly different. \textbf{c}, Diffuse (top) and specular (bottom) reflected colours as a function of $\theta_\text{i}$ and $\rho$. The colours of the diffuse light mainly depend on the spectrum of the individual particle, whereas the colours of the specular light are much richer, due to the thin-film interference experienced by the average field in the particle monolayer. \textbf{d}-\textbf{e}, Rendered images of a macroscopic spherical probe covered by metasurfaces made of spherical particles with different radii, decomposed into the diffuse and specular terms for $f=0.1$. In addition to the global colour change between \textbf{d} and \textbf{e}, note the different greenish and yellowish hues, respectively, of the specular reflection observed in the mirror images of the chessboard bright zones.}
  \label{fig-particle-engineering}
\end{figure*}

Let us now examine the effect of each degree of freedom offered by disordered metasurfaces on visual appearance. We start at the smallest scale with the individual particle resonances, which are expected to drive the dominant colour of metasurfaces~\cite{kristensen2017plasmonic, daqiqeh2020nanophotonic}. We consider the simplest possible metasurface, a monolayer of spherical silicon (Si) particles deposited randomly on a neutral, glass substrate at a surface density $\rho$. Figure~\ref{fig-particle-engineering} summarises our observations and conveys an important conclusion: in general, the colours of the diffuse and specular reflectances differ one from another and may be controlled quasi-independently with disordered metasurfaces.

Figure~\ref{fig-particle-engineering}b shows the diffuse and specular reflectance spectra integrated over the upper hemisphere for unpolarized light incident at an angle $\theta_\text{i}=30^\circ$ and Fig.~\ref{fig-particle-engineering}c shows the corresponding colours predicted by analysing the BRDF data as a function of $\theta_\text{i}$ and $\rho$. The diffuse colours are found to depend weakly on the incident angle and density. They are dominantly driven by the Mie resonances of the individual particles, $\text{d}\sigma_\text{s}/\text{d}\Omega$ in Eq.~\eqref{eq:BRDF-dif}, whose resonance wavelengths shift linearly with the particle size. In contrast, the specular colour is much richer, exhibiting a complex dependence on both $\theta_\text{i}$ and $\rho$, due to the thin-film interference experienced by the average field in a strongly dispersive monolayer on a substrate.

Figure~\ref{fig-particle-engineering}d-e shows rendered images of a spherical probe covered by a metasurface, placed on a curved chessboard and illuminated by directional light. This virtual scene -- a standard in computer graphics -- allows visualizing at once the appearance of the surface at many scattering angles thanks to the spheroid shape, and to distinguish the diffuse and specular components thanks to the contrasted dark and bright areas of the chessboard. Two particle sizes at constant surface coverage $f=\rho \pi r^2$ (i.e., at different densities $\rho$) are considered (additional images for another particle size are reported in the Extended Data Fig.~1). The images clearly evidence the strikingly different contributions from the specular and diffuse components on visual appearance, emphasizing that structural colour is a subtle combination of both. To our knowledge, this simple, albeit important effect has not been previously reported in the fertile literature on structural colours with metasurfaces, which essentially focused on the colour produced by either isolated or periodically-arranged resonant nano-objects in specific illumination and observation conditions~\cite{kristensen2017plasmonic, daqiqeh2020nanophotonic}.

\begin{figure*}[h!]
	\centering
	\includegraphics[width=\textwidth]{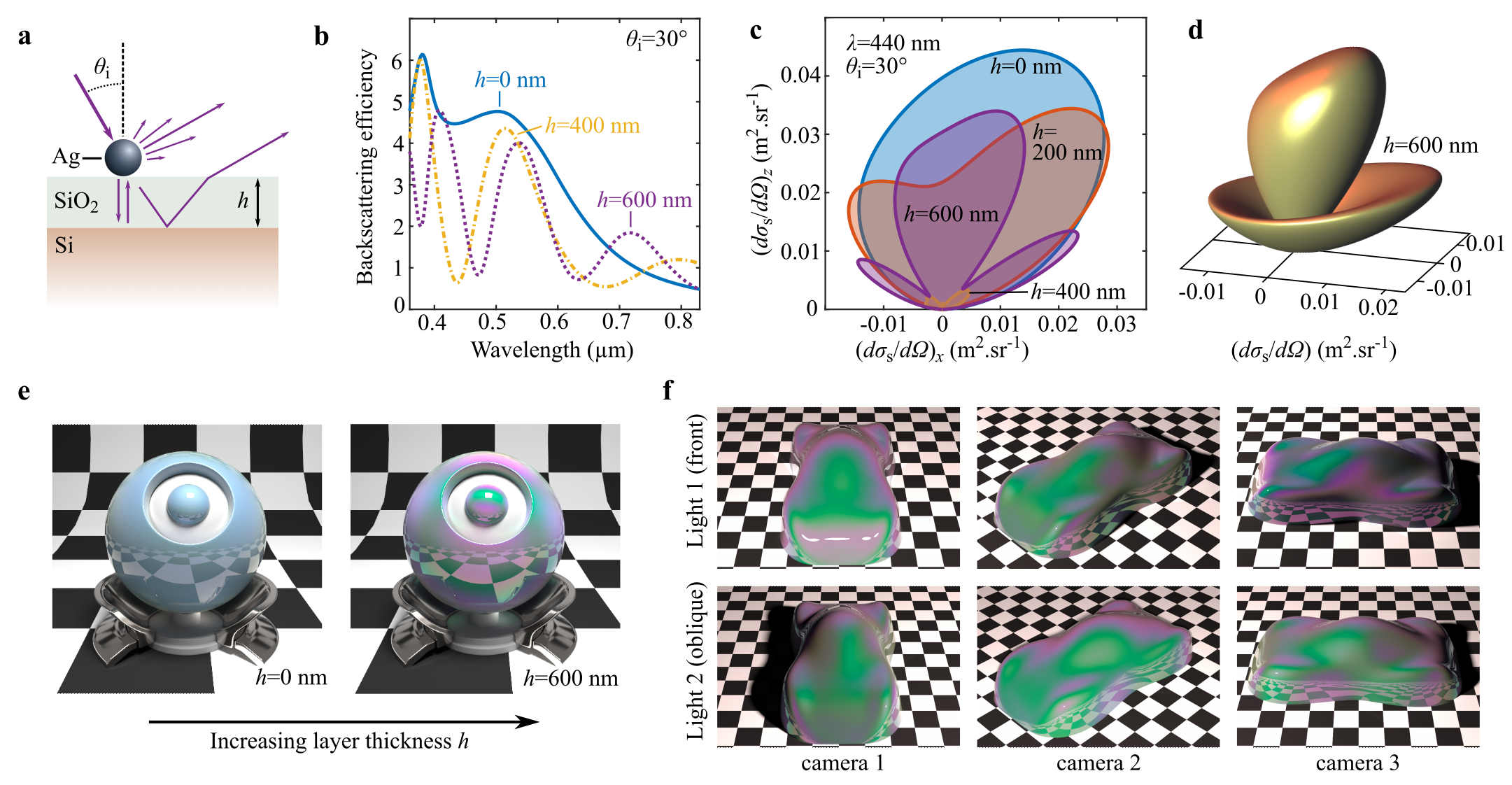}
	\caption{\textbf{Engineering of a layered substrate.} \textbf{a}, Particles are deposited on a layered substrate, composed of a thin dielectric film of thickness $h$ and a semi-infinite film with a different refractive index. Interference between the light scattered by the particle and the light multiply-reflected in the layer create sharp spectral and angular features. \textbf{b}, Backscattering efficiency spectra of an individual Ag particle of radius $r=90$ nm on a SiO$_2$/Si substrate for different values of $h$. \textbf{c,d}, Scattering diagrams of the particle at a wavelength of 400 nm. Angular lobes, whose angular position depend on the wavelength, start to appear with increasing layer thickness $h$. \textbf{e}, Rendered images of metasurfaces made of Ag particles distributed randomly at a surface coverage $f=0.1$ ($\rho = 3.93$ $\mu$m$^{-2}$) as a function of $h$. Diffuse colours emerge with increasing layer thickness due to the modifications of scattering spectra and diagram. \textbf{f}, Rendered images of a speedshape for $h=600$ nm for different viewpoints and light source positions. The images and the Supplementary Video 1 reveal how the colour patterns move with varying viewpoints and illumination conditions. The diffuse iridescence is very different from what would be observed only with classical thin-film interference.}
	\label{fig-interaction-tf-stack}
\end{figure*}

We now proceed with the analysis of the most surprising visual appearances reported in Figs.~\ref{fig1}f-g. The impact of coherent and incoherent scattering on visual appearance can be considerably enriched by considering high-index layered substrates to boost the interaction of light with the particles via multiple scattering~\cite{holsteen2017purcell}. To illustrate this, we consider Ag particles of radius $r=90$ nm deposited randomly on a silica (SiO$_2$) thin film on top of a Si substrate, see Fig.~\ref{fig-interaction-tf-stack}a. As shown in Fig.~\ref{fig-interaction-tf-stack}b-d, changing the SiO$_2$ film thickness $h$ drastically modifies the spectrum and scattering diagram of the scattered light, leading to marked spectral and angular maxima and minima. The number of such features evidently scales with the ratio $nh/\lambda$, where $n$ is the refractive index of the intermediate layer.

Two simulated images of the spherical probe are provided in Fig.~\ref{fig-interaction-tf-stack}e for $h=0$ and $600$ nm (additional images for intermediate thicknesses are reported in the Extended Data Fig.~2). As the thickness increases, vivid diffuse colours appear out of a pale bluish grey. Note that this striking change in appearance is due to spectral and angular variations of the \textit{diffuse} reflectance. Variations of the specularly reflected light responsible for classical thin-film iridescence play a minor role. The peculiarity of this visual effect is yet more evident by looking how the appearance of a non-spherical object changes with varying viewpoints and illumination conditions. Figure~\ref{fig-interaction-tf-stack}e shows rendered images of a ``speedshape'' -- a synthetic object free of sharp edges to avoid very rapid light intensity variations -- under different viewing directions and for two different light source positions. The speedshape predominantly shows lively green and violet diffuse colours, the green preferentially appearing for small incident and viewing angles compared to the surface normal, and the violet for larger angles. Importantly, these coloured patterns dynamically move in a very unusual manner when moving around the object, see the Supplementary Video 1.

Having two or more dominant and markedly different colours moving on the surface of a unique object is an uncommon visual effect, which we may call ``diffuse iridescence''. It cannot be confused with the well-known thin-film iridescence, which is observed in the specular direction only, yielding appearances similar to those of Fig.~\ref{fig-particle-engineering}d-e with the specular term only. Diffuse iridescence is created by both multipeaked spectra and scattering diagrams, which could hardly be reproduced with structuring strategies other than disordered metasurfaces.

Another strategy to introduce spectral and angular features is to implement structural correlations in the disordered pattern. Short-range structural correlations are known to affect significantly light scattering, transport, absorption and emission in disordered systems~\cite{wang2020dependent, vynck2021light}. Disordered metasurfaces with correlated disorder are thus attracting growing attention and recent studies have demonstrated a reduced diffusion around the specular direction for normal incidence~\cite{piechulla2021tailored, sterl2021shaping}. This feature is shown hereafter to persist for all angles of incidence. More notably, the full knowledge of the BRDF (not only at normal incidence) allows us to assess the impact of this scattering property on the appearance of macroscopic curved metasurfaces with correlated disorder and report a stunning visual effect. We further offer a direct observation of this effect on samples large enough to be observable with the naked eye (not through a microscope objective with a high numerical aperture as in~\cite{sterl2021shaping}).

\begin{figure*}[h!]
	\centering
	\includegraphics[width=\textwidth]{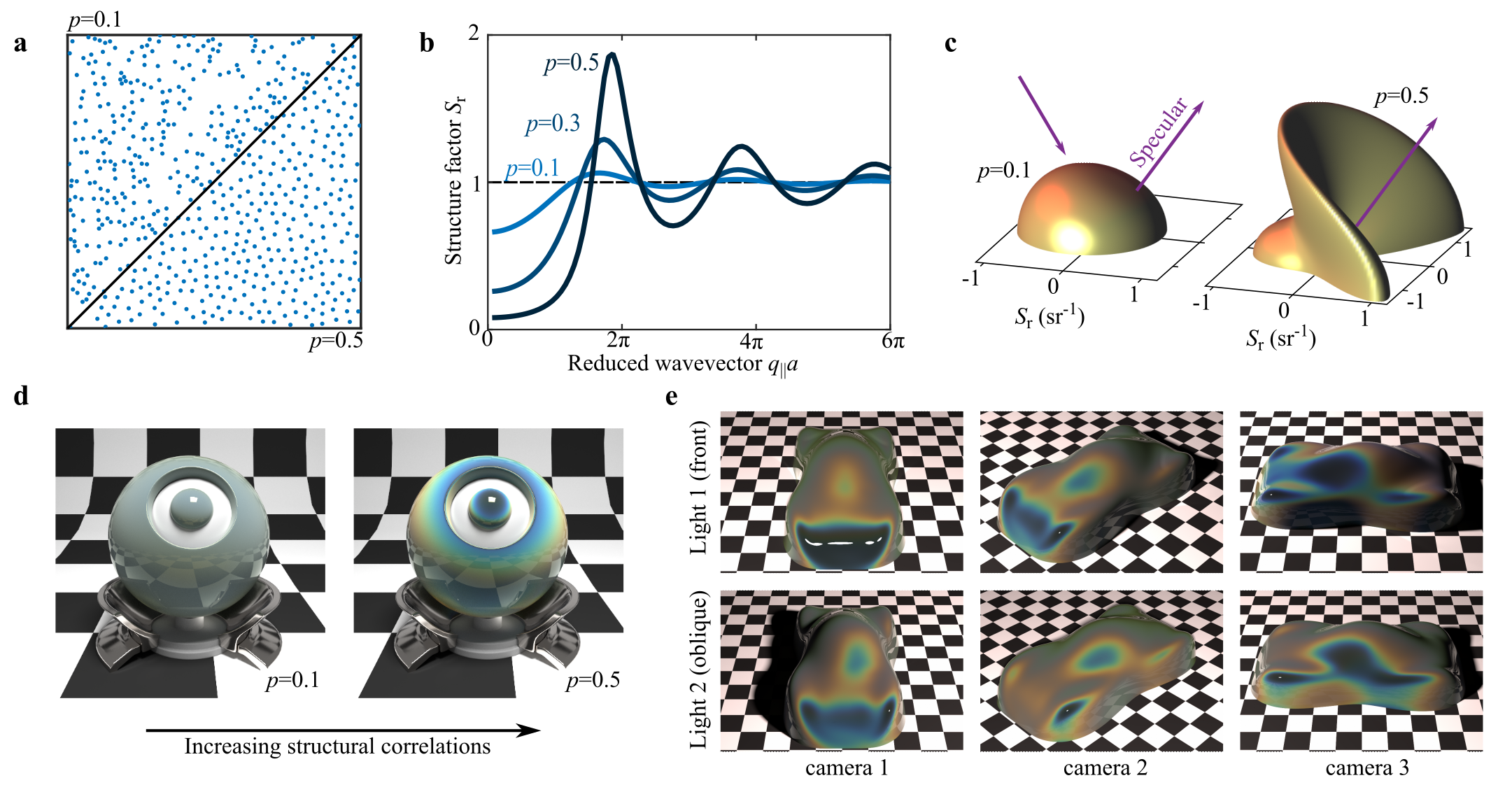}
	\caption{\textbf{Engineering of structural correlations.} \textbf{a}, Examples of spatial point patterns for two different degrees of structural correlations, $p=0.1$ and $0.5$, at a constant particle density. \textbf{b}, Corresponding structure factor $\mathcal{S}_\text{r}$ given as a function of the parallel scattering wavevector $\mathbf{q}_{||}$ and the minimum interparticle distance $a$. Structural correlations can strongly suppress or enhance scattering, depending on the wavelength and the incident angle. \textbf{c}, View of the structure factor in the upper hemisphere with increasing correlation degree $p$ (for $\rho=2.5$ $\mu$m$^{-2}$, $\lambda=450$ nm and $\theta_\text{i}=30^\circ$). Strongly correlated systems exhibit a suppressed diffuse intensity near the specular reflection surrounded by sharp angular lobes. \textbf{d}, Rendered images of the spherical probe for a metasurface composed of Ag particles of radius $r=90$ nm at a density $\rho = 5$ $\mu$m$^{-2}$ on a glass substrate for $p=0.1$ and $0.5$. \textbf{e}, Rendered images of the speedshape in the case of a strongly correlated system ($p=0.5$) for different viewpoints and light source positions. In the diffuse halo, the suppression of the diffuse reflectance and the formation of vivid surrounding colours are following the specular direction, contrary to the diffuse iridescence reported in Fig.~\ref{fig-interaction-tf-stack}e, see also the Supplementary Video 1.}
	\label{fig-struct-corr-Ag}
\end{figure*}

A simple way to control structural correlations consists in imposing a minimum interparticle distance $a$ between particles, leading to an effective surface coverage (or ``packing'' fraction) $p=\rho \pi a^2 /4$, see Fig.~\ref{fig-struct-corr-Ag}a. The impact of structural correlations on the diffuse reflectance can be apprehended by analysing the structure factor $\mathcal{S}_\text{r}$ in Eq.~(\ref{eq:BRDF-dif}), obtained from the Fourier transform of the correlation function between pairs of particles. The structure factor essentially renormalises the scattering diagram by individual particles by taking far-field interference into account. Figure~\ref{fig-struct-corr-Ag}b shows the structure factors of two point patterns with different packing fractions $p$ as a function of the in-plane scattering wavevector $q_{\parallel} = |\mathbf{q}_{\parallel}|$ with $\mathbf{q}_{\parallel}=\mathbf{k}_{\text{i},\parallel}-\mathbf{k}_{\text{s},\parallel}$. Upon increasing $p$, $\mathcal{S}_\text{r}$ is substantially reduced around the specular direction ($q_{\parallel} \sim 0$) and significantly enhanced for $q_\parallel \sim 2\pi/a$. Structural correlations considerably impact the radiation diagram of the diffuse light. It significantly suppresses scattering around the specular direction and, very importantly also for visual appearance, sharply enhances it at farther angles, see Fig.~\ref{fig-struct-corr-Ag}c. This interesting property predicted from structure-factor arguments simply relying on the statistical position of the particles is fully validated by full-wave computational results obtained for a large (but finite) monolayer of spherical nanoparticles as evidenced in the Supplementary Note 2. By analogy with the previous effect, the angular position and number of lobes are driven here by the ratio $a/\lambda$.

Figure~\ref{fig-struct-corr-Ag}d shows rendered images of the spherical probe covered by metasurfaces composed of Ag particles on a glass substrate at density $\rho = 5$ $\mu$m$^{-2}$ for two different correlation degrees $p=0.1$ and $p=0.5$ (additional images for other $p$ and $\rho$ are reported in the Extended Data Fig.~3, and for metasurfaces made of Si particles in the Extended Data Fig.~4). Short-range structural correlations lead to a strong and broadband attenuation of the scattered intensity around the specular direction (the dark bluish isotropic area on the top) surrounded by a rainbow-like colour gradient at larger angles due to the first peak of $\mathcal{S}_\text{r}$. The object appearance is dominantly specular in areas where the diffuse scattering is reduced. The extent of the darkened area can be simply controlled by changing the particle density, as shown in the Extended Data Figs. 3 and 4. As evidenced with the rendered images of the speedshape reported in Fig.~\ref{fig-struct-corr-Ag}e and in the Supplementary Video 1, this visual effect, which may be coined as a ``diffuse halo'', is completely different from the diffuse iridescence presented in Fig.~\ref{fig-interaction-tf-stack}: the colour pattern now always strictly follows the specular direction. This explains the stunning appearance first reported in Fig.~\ref{fig1}f, where light incident on the car essentially comes from a strip of diffuse sky. In fact, the tight relation with the incident light direction and position implies that a macroscopic object covered by a such a metasurface will acquire a remarkably rich variety of appearances depending on the lighting environment. This richness is shown in the Supplementary Video 2, where the speedshape is placed under either a diffuse or a clear sky, and in the Extended Data Figs. 5 and 6, where we simulate the visual appearance of a car in realistic outdoor and indoor environments.

%%%%%%%%%%%%%%%%%%% EXPERIMENTS %%%%%%%%%%%%%%%%%%%

\section*{Experimental proof of concept}

We conclude our study by the experimental demonstration of the diffuse-halo effect on centimetre-scale metasurfaces. By contrast with a recent study on correlated disordered metasurfaces~\cite{sterl2021shaping}, we consider here the true visual appearance of a macroscopic surface as observed by the naked eye and study how it evolves with the illumination and viewing conditions. As explained above, these dynamical properties are essential for perception.

\begin{figure*}[h!]
	\centering
	\includegraphics[width=175.740mm]{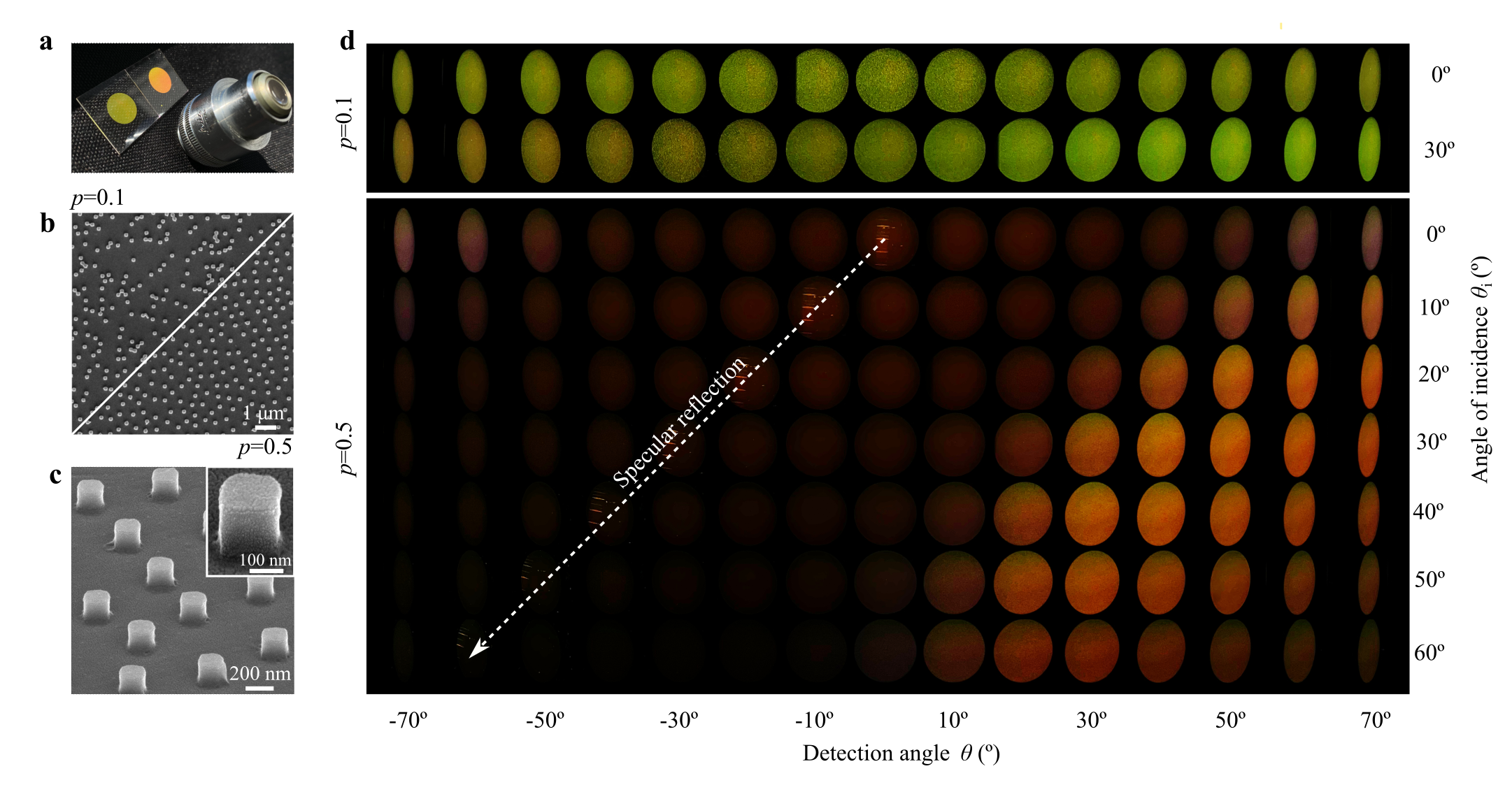}
	\caption{\textbf{Experimental demonstration of the diffuse halo due to short-range structural correlations}. \textbf{a}, Photograph of the fabricated centimetre-scale metasurfaces next to a microscope objective. The dissimilarity in the diffuse colour is due to the ageing of the negative resist, used for the fabrication of the two samples, over an interval of two months. \textbf{b},\textbf{c}, Corresponding scanning electron microscope (SEM) micrographs. In \textbf{c}, the SEM image is tilted by $57^\circ$. \textbf{d}, Sequence of photographs of the metasurfaces at different illumination and viewing angles. The light source is the collimated beam of a solar simulator. As predicted, the weakly-correlated metasurface ($p=0.1$, top) exhibits a nearly-uniform diffuse colour independent of the illumination angle, whereas the strongly-correlated metasurface ($p=0.5$, bottom) leads to a diffuse halo, characterized by darker and brighter regions that follow the direction of specular reflection (white arrow). The complete recording of the experiment for the series $\theta_\text{i} = 30^{\circ}$ is given in the Supplementary Video 3. Quantitative BRDF measurements realized on the two metasurface samples and reported in the Supplementary Note 5 support our conclusions.}
	\label{Fig:5}
\end{figure*}

Figure~\ref{Fig:5}a shows a photograph of the two samples fabricated by top-down lithography techniques (see Methods). The metasurfaces are composed of Si nanodisks with a nominal $175 \times 175~{\rm nm^2}$ square base and a height of $145~{\rm nm}$ on a SiO$_2$ substrate, using the same density, $\rho=3.14~{\rm \mu m^{-2}}$, in both cases but distinct packing fractions, $p = 0.1$ and $p = 0.5$. Scanning electron microscope (SEM) images of the samples are shown in Fig.~\ref{Fig:5}b-c. Owing to their macroscopic flatness, the two samples exhibit a uniform, coloured and glossy appearance. The appearance of the strongly correlated structure ($p=0.5$) yet changes quite drastically in brightness when moving the sample orientation or the viewpoint. To be more quantitative, we perform a series of photographs of the samples with a calibrated camera of a smartphone mounted on a spectrogoniometer, as a function of both incident and detection angles, see Fig.~\ref{Fig:5}d. The samples are illuminated by the collimated beam of a solar simulator and the camera is configured as to produce photographs that faithfully reproduce the appearance of the samples as observed by naked eye in similar conditions. The weakly-correlated metasurface (top) keeps a relatively constant luminance -- similarly to a Lambertian surface -- with a slight change of colour with increasing detection angle and a weak dependence on the incident angle. In sharp contrast, the strongly-correlated metasurface (bottom) fades to black as the camera approaches the direction of specular reflection and appears brighter at larger angles. More strikingly, these darker and brighter regions follow the angle of specular reflection when the incident angle is varied. In a different frame, this 2D map describes the appearance of a \textit{curved} surface under a fixed illumination and a moving viewpoint (in a specific plane of incidence). This corroborates the visual effect observed at grazing incident angles on the speedshape in the Supplementary Video 1, namely a bright diffuse colour in backreflection and an almost purely specular, black appearance on the opposite side. These experiments provide strong evidence that the engineering of disordered metasurfaces yield novel, exotic visual effects.

%%%%%%%%%%%%%%%%%%% DISCUSSION %%%%%%%%%%%%%%%%%%%

\section*{Outlook}

Visual appearance is a much broader and more powerful concept than colour. Our results constitute a fundamental advance on how scattering at the nanoscale and mesoscale may be used to control visual appearance at the macroscale. Engineering the interaction between particles via a layered substrate or correlated disorder leads to a myriad of different specular and diffuse reflectances, and enhances dynamical effects perceived when the viewing or illumination directions change, with no equivalent in traditional materials. Other stunning visual appearances may be achieved by further exploiting photonic or plasmonic resonances of non-spherical particles, strong interaction regimes with metallo-dielectric substrates~\cite{akselrod2015} or different types of structural correlations~\cite{florescu2009designer, salvalaglio2020hyperuniform}. These concepts can further be generalized to transmissive metasurfaces, that may act as novel coatings on windows and displays to provide new functionalities for augmented reality applications~\cite{hsu2014transparent}.

We used top-down nanofabrication to experimentally demonstrate a stunning visual effect with centimetre-scale metasurfaces. Our full-wave simulations reported in Supplementary Note 3 constitute strong evidence that the two visual effects reported in Figs.~\ref{fig-interaction-tf-stack} and \ref{fig-struct-corr-Ag} with plasmonic particles can easily withstand particle size polydispersity of 10 \%. Thus, we envision that these effects could be observed with large-area metasurfaces fabricated by bottom-up (e.g., colloidal) approaches, possibly on curved substrates~\cite{akselrod2015}, which is very encouraging for future developments.

The visual appearance of disordered metasurfaces is difficult to anticipate in general, considering the large variabilities in nanoscale and mesoscale phenomena, object shapes and lighting environment, even when the BRDF is known. Enhanced by predictive rendering, our study allowed us to establish recipes of stunning visual effects with disordered metasurfaces. A formidable goal for future work will be to enable inverse design strategy~\cite{weyrich2009fabricating, levin2013fabricating, andkjaer2014inverse, auzinger2018computational} to determine realistic metasurface parameters yielding targeted visual appearances in specific lighting conditions -- a challenge that may only be achieved by merging concepts in nanophotonics and computer graphics.

%%%%%%%%%%%%%%%%%%% ADDENDUM %%%%%%%%%%%%%%%%%%%

\section*{Acknowledgements}
All authors are grateful to Pascal Barla (INRIA Bordeaux Sud-Ouest, Talence, France) for very stimulating and fruitful discussions on the BRDF model and the interpretation of visual appearances. Pascal Barla declined being an author of the present work for ecological reasons. P.L. acknowledges Franck Carcenac (LAAS, Toulouse, France) for his attention and diligence in fabricating the metasurfaces under the RENATECH program of CNRS. X.G. and P.L. acknowledge Philippe Bouyer (LP2N, Talence, France) for inspiring discussions at the initial stage of the project. K.V. and P.L. acknowledge Jean-Paul Hugonin (LCF, Palaiseau, France) for his help in the development of the full-wave simulation tool used to test the model accuracy. P.L. thanks Louise-Eug\'{e}nie Bataille, Philippe Teulat, Arnaud Tizon and Louis Bellando for their help in developing the goniospectrometer setup. P.L. and A.A. acknowledge Jacques Leng (LOF, Pessac, France) for giving free access to the solar simulator and Bertrand Simon (LP2N, Talence, France) for fruitful discussions on the experimental measurements. This work has received financial support from the French State and the R\'egion Nouvelle-Aquitaine under the CPER project ``CANERIIP'', from CNRS through the MITI interdisciplinary programs, and from the french National Agency for Research (ANR) under the projects ``NanoMiX'' (ANR-16-CE30-0008), ``VIDA'' (ANR-17-CE23-0017), and ``NANO-APPEARANCE'' (ANR-19-CE09-0014).

\section*{Author Contributions}
K.V. elaborated the BRDF model with feedbacks from R.P. and P.L., performed the electromagnetic calculations and compiled the numerical BRDF data. R.P. and A.D. developed and used the rendering tools to obtain the appearance of nanostructured objects. K.V. and P.L. developed the full-wave simulation tool used to test the BRDF model accuracy. A.A. and P.L. developed the experimental setups. A.A. performed the experimental measurements and calibrated photographs. All authors discussed the results and their interpretation, and contributed to writing the manuscript.

\section*{Competing Interests}
The authors declare the following competing interests: Patent deposited on the control of visual appearance with disordered metasurfaces [Applicants: Universit\'{e} de Bordeaux, Centre National de la Recherche Scientifique (CNRS), Institut d'Optique Th\'{e}orique et Appliqu\'{e}e, and Universit\'{e} Paris-Saclay; Inventors: K.V., R.P., X.G. and P.L.; Filing date: February 1st, 2021; Application number: FR 2100948]. The remaining authors declare no competing interests.

\newpage

%%%%%%%%%%%%%%%%%%% BIBLIOGRAPHY %%%%%%%%%%%%%%%%%%%

%\bibliographystyle{naturemag}
%\bibliography{library}

\newpage

%%%%%%%%%%%%%%%%%%% METHODS %%%%%%%%%%%%%%%%%%%

\section*{Methods}

\subsection*{BRDF model}

The BRDF $f_\text{r}$ is a radiometric quantity~\cite{nicodemus1977geometrical} describing how an incident planewave at angles $\theta_\text{i}$ and $\phi_\text{i}$ is scattered by a planar surface into outgoing planewaves with angles $\theta_\text{s}$ and $\phi_\text{s}$. It is conveniently composed of a specular term and a diffuse term, $f_\text{r} = f_\text{r}^\text{spe} + f_\text{r}^\text{dif}$. As shown in the Supplementary Note 1, the specular term $f_\text{r}^\text{spe}$ is driven by the reflection of the average field and, for a statistically translationally-invariant surface, can be expressed as
\begin{equation}\label{eq:BRDF-spe}
f_\text{r}^\text{spe}(\mathbf{k}_\text{s},\hat{\bm e}_\text{s},\mathbf{k}_\text{i},\hat{\bm e}_\text{i}) = \frac{\delta(\pi-\theta_\text{s} - \theta_\text{i}) \delta(\phi_\text{s} - \phi_\text{i}) \delta_{\hat{\bm e}_\text{s},\hat{\bm e}_\text{i}}}{\sin \theta_\text{s} \cos \theta_\text{s}} |r_\text{st} (\mathbf{k}_\text{s},\hat{\bm e}_\text{s})|^2,
\end{equation}
where $\delta(\cdot)$ and $\delta_{ij}$ are the Dirac delta and Kronecker delta functions, $\mathbf{k}_\text{i}$, $\mathbf{k}_\text{s}$, $\hat{\bm e}_\text{i}$ ,$\hat{\bm e}_\text{s}$ are the wavevectors and polarisations of the incident and scattered planewaves, and $r_\text{st}$ is the complex reflection coefficient of the particle monolayer on the substrate. We use the state-of-the-art multiple-scattering model developed in~\cite{garcia2012multiple} to calculate $r_\text{st}$.

The diffuse term $f_\text{r}^\text{dif}$, instead, is given by the fluctuating field. In the Supplementary Note 1, we show that a faithful approximate expression for $f_\text{r}^\text{dif}$ is
\begin{equation}\label{eq:BRDF-dif}
f_\text{r}^\text{dif}(\mathbf{k}_\text{s},\hat{\bm e}_\text{s},\mathbf{k}_\text{i},\hat{\bm e}_\text{i}) = \rho \frac{\text{d}\sigma_\text{s}}{\text{d}\Omega} (\mathbf{k}_\text{s},\hat{\bm e}_\text{s},\mathbf{k}_\text{i},\hat{\bm e}_\text{i}) \mathcal{S}_\text{r}(\mathbf{k}_\text{s},\mathbf{k}_\text{i}) \frac{\mathcal{C}(\mathbf{k}_\text{s},\hat{\bm e}_\text{s},\mathbf{k}_\text{i},\hat{\bm e}_\text{i})}{\cos \theta_\text{i} \cos \theta_\text{s}}.
\end{equation}
$f_\text{r}^\text{dif}$ is proportional to the particle density $\rho$, as expected for a method asymptotically exact for sparse systems. $\text{d}\sigma_\text{s}/\text{d}\Omega$ is the scattering diagram of an individual particle; it is numerically computed with an in-house freeware code~\cite{Yang2016} that takes into account the interaction of the particle with the substrate. The structure factor $\mathcal{S}_\text{r}$ that describes spatial correlations between pairs of particles is computed with the semi-analytical Baus-Colot model for hard-disk liquids~\cite{baus1987thermodynamics}. Finally, $\mathcal{C}$ is a correction term retaining all our attention, as it considerably increases the model accuracy. It ensures that $f_\text{r}^\text{dif}$ always vanishes as $\theta_i$ or $\theta_s$ tend towards $\pi/2$, whatever the polarization or wavelength. It further ensures that the model remains accurate even at relatively large filling fractions (up to $f \sim 0.15-0.2$) and guarantees the reciprocity of the BRDF as was numerically verified with an accuracy of $10^{-6}$. We propose here a heuristic model to determine $\mathcal{C}$. Its expression is derived in the Supplementary Note 1. Supplementary Note 2 reports some of our numerical tests validating the predictive force of the model by comparison with multiple-scattering computations on large and dense monolayers of particles using a dedicated in-house full-wave computational method~\cite{langlais2014cooperative, jouanin2016designer, bertrand2020global}. This computational method is also used in the Supplementary Note 3, where we show that the exotic scattering properties reported in Figs.~\ref{fig-interaction-tf-stack} and \ref{fig-struct-corr-Ag} withstand particle size polydispersity up to 10 \%, at least.

The BRDF of a disordered metasurface, given by Eqs.~(\ref{eq:BRDF-spe}) and (\ref{eq:BRDF-dif}), requires computing the scattering amplitude resolved in frequency, angle and polarization and the absorption cross-section of an individual particle, both in free space and on a layered medium, for each possible excitation (wavelength, incident angle and polarization). The calculations were performed for both $s$- and $p$-polarisations, spanning the wavelength range [360-830] nm uniformly in steps of 10 nm, and for incident and scattered angles, $\theta_\text{i}$ and $\theta_\text{s}$ ranging from 0 to 89.5$^\circ$ in steps of 0.5$^\circ$, and $\delta\phi=\phi_\text{s}-\phi_\text{i}$ from 0 to 360$^\circ$ in steps of 1$^\circ$. We have verified that the angular and spectral resolutions were sufficient to yield converged rendered images, see the validation tests in the Supplementary Note 4.

The material permittivities for SiO$_2$, Si and Ag are taken from tabulated data in Refs.~\cite{malitson1965interspecimen}, \cite{green1995optical} and \cite{johnson1972optical}, respectively and we used $n=1.5$ for glass.

\subsection*{Spectral rendering}

To generate a synthetic image, one needs to simulate the light transport inside a chosen virtual scene observed from a virtual viewpoint (i.e., a virtual camera)~\cite{pharr2016physically}. A virtual scene is obtained by modelling (or measuring from the real-world) the shapes of the objects and the light sources, the materials reflectance (BRDF) and transmittance (BTDF), and the spectral emittance of the light sources. The radiometric quantities (BRDF, BTDF and emittance) have angular, spatial and spectral dependencies but also vary with the polarisation. For example, the scene in Fig.~\ref{fig1}d-g consists of a car model (freely available on Blend Swap~\cite{blendswapM3} and also used in the Extended Data Figs. 5 and 6) and an environment map (here, the Uffizi gallery~\cite{uffizi}) that approximates the environment by an image where each pixel represents a distant light source in a given direction. The environment maps of the outdoor rendered images in Fig.~\ref{fig1}, Extended Data Fig.~5 and Supplementary Video~2 come from~\cite{uffizi} and Poly Haven~\cite{hdri}. The environment maps were converted to spectral images by using a classical optimization procedure (quadratic optimization with smoothness and positivity constraints) similarly to Ref.~\cite{meng2015physically}. This guarantees that the final spectrum of every individual pixel, once projected back into the sRGB space, gives the same color as the original one.

Regarding the materials, we assigned the BRDF model to the car body, whereas the windshield material was modeled using Fresnel equations with an index of refraction of 1.45 for the whole glass. The tire material was assumed Lambertian whereas the rims were modeled using a microfacet-based BRDF with sharp angular response for its distribution (GGX)~\cite{walter2007microfacet}, and a measured index of refraction corresponding to chromium~\cite{johnson1974optical}.

Simulating the light transport is done by solving the recursive rendering equation~\cite{kajiya1986rendering}. This equation states that the equilibrium radiance (in W.m$^{-2}$.sr$^{-1}$ per wavelength) leaving a point of a surface is the sum of emitted and reflected radiance under a geometric optics approximation. The rendering equation is therefore directly related to the law of conservation of energy.

We solved the rendering equation with Monte-Carlo computations~\cite{kalos2009monte}. In the context of computer graphics, a Monte-Carlo sample is a geometric ray carrying radiance along its path, which is stochastically constructed (e.g., using Russian roulette)~\cite{pharr2016physically}. To speedup the convergence, we used multiple importance sampling techniques~\cite{pharr2016physically} to favour paths that carry most energy (thus reducing the variance) without introducing  bias in the computation. Finally, we considered time-invariant and polarisation-averaged radiometric quantities when solving the rendering equation per wavelength.

The output of the light transport simulation is a synthetic image where each pixel stores the spectral distribution of the radiance flowing through it. The spectral synthetic images were obtained by first projecting them to the CIE-XYZ color space~\cite{wyszecki1969color} and then converting them to sRGB color space using standard procedure~\cite{sRGB1999}. Thus, to faithfully visualise the rendered images reported here, the monitor should be calibrated to sRGB.

All presented images were generated with our own spectral rendering engine~\cite{mrfweb} based on Optix~\cite{parker2010optix}. We checked the validity of our predictions by comparing rendering results obtained using a closed-source spectral rendering engine (Nvidia iRay~\cite{keller2017iray}). Nevertheless, our spectral rendering engine allows a complete control of the rendering pipeline and easier integration of new BRDF models.

The time to generate a synthetic image depends on the image resolution (number of pixels), the number of paths per pixel as well as the number of wavelengths (46 in our case). As an indication, each image of Fig.~\ref{fig1} has a full-HD resolution ($1920 \times 1080$ pixels) and was rendered with 6000 paths per pixel in 15 minutes using a 32-cores Intel(R) Xeon(R) Gold 6142 CPU @ 2.60GHz with two Nvidia TESLA V100 GPUs.

\subsection*{Design and fabrication of the metasurface samples}

The metasurfaces were designed using a Poisson disk sampling algorithm~\cite{bridson2007fast} available at~\cite{patel} to generate the nanodisk coordinates, using as input parameters the sample size, particle density $\rho$ and minimum interparticle distance $a$, which determine the surface coverage and packing fraction $f$ and $p$, respectively, with $p \geqslant f$. The generated files containing the coordinates ($x,y$) were then converted to GDS files for the electron-beam lithography.

The metasurfaces were fabricated in the LAAS nanofabrication facilities by a combination of standard electron beam lithography and reactive ion etching. A $\sim 150~{\rm nm}$ polycrystalline silicon layer was deposited on a glass coverslip. Negative (Micro Resist Technology, ma-N 2405) and conductive (Allresist GmbH, AR-PC 5090.02) resists were spin-coated and exposed to an electron beam with $20~{\rm kV}$ acceleration voltage and $300~{\rm pA}$ current. The negative resist was developed for $50~{\rm s}$ in a solution of MF CD-26. The process was followed by fluorine-based reactive ion etching to finally produce the silicon nanodisks and oxygen plasma etching to remove the remaining negative resist.

\subsection*{Goniospectrometer setup}

The BRDF measurements reported in the Supplementary Note 5 were performed with a homemade spectrophotometer system equipped with two stepper motor rotation stages (Newport, URS75 and URS150) that can vary the angles of incidence $\theta_i$ and detection $\theta$. A supercontinuum laser (Leukos, Rock 400) was used as a broadband source for the BRDF measurements. The infrared wavelengths of the laser beam were rejected with a short-pass filter (Schott KG-1). Just after, the beam was expanded with a telescopic system and passed through a diaphragm to control the beam size ($1~{\rm cm}$). An optical fiber mounted on the motorized stage and connected to a spectrometer (Ocean Insight HDX-XR) resolved the scattered light spectrally. A LabView user interface allowed the automation of the detector and sample movement and the subsequent recording of the corresponding spectra.

\subsection*{Photographs setup}

The series of photographs in Fig.~\ref{Fig:5}d was taken with the camera of an iPhone 11 (12 MP Wide camera, f/1.8 aperture). The exposure time as well as the focusing distance were blocked in the camera ensuring that the detection conditions were fixed during the complete photo shooting. The iPhone was mounted on the goniometric setup, with the smartphone lens located slightly above the plane of incidence to avoid the specular reflection and at about $12~{\rm cm}$ away from the samples. Since the entrance pupil size of the human eye, typically between 2-5 mm, is close to the entrance pupil diameter of the camera, the photographs are recorded with a numerical aperture comparable to those reached by the normal eye in similar conditions. Therefore, the photograph series faithfully reproduce the visual appearance of the sample. The metasurfaces were illuminated with the collimated beam of a solar simulator (ASAHI SPECTRA, HAL-320). The intensity of the solar simulator was set to ensure that the camera captures the appearance of the metasurface and was kept constant during the photo shooting. Similar settings were used for the Supplementary Video 3.

\section*{Data availability}

The datasets underlying the figures of the current study are available either in the Zenodo repository, https://doi.org/10.5281/zenodo.6327176, or from the corresponding authors upon reasonable request.

\section*{Code availability}

The codes used to compute the scattering diagrams of particles and to generate the synthetic images in this study are publicly available at https://doi.org/10.5281/zenodo.3609149 and https://mrf-devteam.gitlab.io/mrf/main.md.html, respectively. Additional codes used in this study are available from the corresponding authors upon reasonable request.

%%%%%%%%%%%%%%%%%%% BIBLIOGRAPHY %%%%%%%%%%%%%%%%%%%

\end{document}